\documentclass[11pt,a4paper,twoside,groupcitations]{article}
\usepackage[]{latexsym,amsmath,amssymb,upgreek}
\usepackage[]{epsfig}
\usepackage{color}
\usepackage{braket}
\usepackage{graphicx}
\usepackage{bm}
\usepackage{cite}
\newcommand{\be}{\begin{eqnarray}}
\newcommand{\ee}{\end{eqnarray}}
\def\beq{\begin{equation}}
\def\eeq{\end{equation}}

\newcommand{\pro}[2]{\mbox{$\langle\, #1 \mid #2\,\rangle$}}
\newcommand{\expec}[1]{\mbox{$\langle\, #1\,\rangle$}}

\renewcommand{\a}{\hat a}
\newcommand{\ac}{\hat a^{\dagger}}

\newcommand{\lp}{\ell_{\rm p}}
\newcommand{\mpl}{m_{\rm p}}
\newcommand{\gn}{G_{\rm N}}
\newcommand{\rh}{r_{\rm H}}
\newcommand{\Rh}{R_{\rm H}}
\newcommand{\dd}{\mbox{${\rm d}$}}

\title{\bf Horizon quantum mechanics for coherent quantum black holes}
\author{Wenbin~Feng$^{ab}$\thanks{E-mail: wenbin.feng2@unibo.it},
$\ $
Andrea~Giusti$^{cd}$\thanks{E-mail: A.Giusti@sussex.ac.uk},
$\ $
and
Roberto~Casadio$^{abd}$\thanks{E-mail: casadio@bo.infn.it}
\\
\\
$^a${\em Dipartimento di Fisica e Astronomia, Universit\`a di Bologna}
\\
{\em via Irnerio~46, 40126 Bologna, Italy}
\\
\\
$^b${\em I.N.F.N., Sezione di Bologna, I.S.~FLAG}
\\
{\em viale B.~Pichat~6/2, 40127 Bologna, Italy}
\\
\\
$^c${\em Department of Physics and Astronomy}
\\
{\em University of Sussex, Brighton, BN1 9QH, United Kingdom}
\\
\\
$^d${\em Alma Mater Research Center on Applied Mathematics - AM$^2$}
\\
{\em Via Saragozza 8, 40123 Bologna, Italy}
}
\date{}
\begin{document}
\maketitle
\begin{abstract}
The formalism of the horizon quantum mechanics is applied to electrically neutral and spherically symmetric
black hole geometries emerging from coherent quantum states of gravity to compute the probability
that the matter source is inside the horizon.
We find that quantum corrections to the classical horizon radius become significant if the matter core
has a size comparable to the Compton length of the constituents and the system is indeed a black hole with
probability very close to one unless the core radius is close to the (classical) gravitational radius.
\end{abstract}
\section{Introduction}
\label{S:intro}
\setcounter{equation}{0}
Coherent quantum states can be employed to describe the static and spherically
symmetric Schwarz\-schild black holes as emergent (semi)classical
geometries~\cite{Casadio:2021eio}.~\footnote{For
studies of their thermodynamics and configurational entropy, see
Refs.~\cite{Casadio:2023pmh,Feng:2024bsx,Belfiglio:2024qsa} and, for more
phenomenological consequences, see Ref.~\cite{Urmanov:2024qai}.}
Such a construction can be straightforwardly extended to the Reissner-Nordstr\"om
black holes~\cite{Casadio:2022ndh,Neznamov:2023hpi} and the semiclassical
approximation for rotating geometries can also be obtained from spherically symmetric
cases~\cite{Casadio:2023iqt,Contreras:2021yxe,Neznamov:2024zll}.
It is important to remark that this approach implies the removal of the central singularities
by the presence of a quantum matter core that could therefore lead to phenomenological
signatures of the kinds analysed in Ref.~\cite{Arrechea:2024nlp}.
\par
The presence of horizons in the above approach can only be established from
semiclassical arguments, that is by considering the quantum corrected metric
\be
\dd s^2
=
-f(r)\,\dd t^2
+
h(r)\,\dd r^2
+
r^2\,\dd\Omega^2
\ ,
\label{metric}
\ee
where $\dd\Omega^2=\dd\theta^2+\sin^2\theta\,\dd\phi^2$ and
\be
f
=
h^{-1}
=
1+2\,V_{\rm q}(r)
\ .
\label{Vq}
\ee
In the above, the function $V_{\rm q}=\bra{V}\hat V(r)\ket{V}$ is the expectation value
of the relevant metric field on the coherent quantum state $\ket{V}$.
The locations of horizons are then given by solutions $r=\rh$ of the classical
equation $f(r)=0$.
We remark that the Kerr-Schild form~\eqref{metric}~\cite{Kerr:1965wfc} in which the metric
has components $f=h^{-1}$ was chosen because it includes all known spherically
symmetric black holes in General Relativity.
However, the presence of an event horizon only requires $f\,h=1$ at $r=\rh$,
so that one could also consider recovering the semiclassical geometry from a more
general coherent quantum state $\ket{f,h}$ such that $f=\bra{f,h}\hat f\ket{f,h}$ and
$h=\bra{f,h}\hat h\ket{f,h}$.
Such a generalisation is left for future developments.
\par
The horizon quantum mechanics was introduced in Refs.~\cite{Casadio:2013tma,Casadio:2013aua}
(see also Ref.~\cite{Casadio:2015qaq} for a review)
to compute the probability of the presence of horizons associated with static and spherically
symmetric matter sources in a given quantum state $\ket{\psi_{\rm S}}$.
We recall that the Einstein field equations for a source of energy density $\rho=\rho(r)$ imply that~\footnote{We
shall use units with $c=1$, the Newton constant $\gn=\lp/\mpl$ and the Planck constant $\hbar=\lp\,\mpl$,
where $\lp$ and $\mpl$ are the Planck length and mass, respectively.}
\be
f
=
1-\frac{2\,\gn\,m(r)}{r}
\ ,
\ee
where the Misner-Sharp-Hernandez mass function~\cite{Misner:1964je,Hernandez:1966zia}
is given by
\be
m(r)
=
4\,\pi\int_0^r \rho(x)\,x^2\,\dd x
\ .
\label{M}
\ee
An horizon then exists if there are values of $r=\rh$ such that $2\,\gn\,m(\rh)=\rh$.
A quantum mechanical description is obtained by replacing the classical energy density
with the energy decomposition of the source wavefunction,
\be
\ket{\psi_{\rm S}}
=
\sum_E\,C(E)\ket{\psi_E}
\ ,
\ee
where the sum represents the spectral decomposition in Hamiltonian eigenmodes,
\be
\hat H\ket{\psi_E}
=
E\ket{\psi_E}
\ ,
\ee
and $H$ will depend on the model we wish to consider.
Upon expressing $E$ in terms of the gravitational Schwarzschild radius,~\footnote{For the local version
of the formalism, see Ref.~\cite{Casadio:2016fev}.}
$E={\rh}/{2\,\gn}$, we obtain the horizon wavefunction
\be
\psi_{\rm H}(\rh)
\equiv
\pro{\rh}{\psi_{\rm H}}
=
\mathcal{N}_{\rm H}
\sum_{E={\rh}/{2\,\gn}}\,C(E)
\ ,
\ee
whose normalisation $\mathcal{N}_{\rm H}$ is fixed in the Schr\"odinger scalar product
\be
\pro{\psi_{\rm H}}{\phi_{\rm H}}
=
4\,\pi
\int_0^\infty
\psi_{\rm H}^*(\rh)\,\phi_{\rm H}(\rh)\,\rh^2\,\dd \rh
\ .
\ee
\par
The normalised wavefunction yields the probability density
for the values of the gravitational radius $\rh$ associated with the source
in the quantum state $\ket{\psi_{\rm S}}$, namely
\be
\mathcal P_{\rm H}(\rh)
=
4\,\pi\,\rh^2\,|\psi_{\rm H}(\rh)|^2
\ .
\label{Phrh}
\ee
Moreover, the probability density that the source lies inside its own gravitational radius
will be given by
\be
{\mathcal P}_<(\rh)
=
P_{\rm S}(\rh)\,{\mathcal P}_{\rm H}(\rh)
\ ,
\label{PrlessH}
\ee
where
\be
P_{\rm S}(\rh)
=
4\,\pi\,\int_0^{\rh}
|\psi_{\rm S}(r)|^2\,r^2\,\dd r
\ee
is the probability that the source is found inside a sphere of radius $r=\rh$.
Finally, the probability that the object described by the state $\ket{\psi_{\rm S}}$ is a
black hole will be obtained by integrating Eq.~\eqref{PrlessH} over all possible
values of the gravitational radius, namely
\be
P_{\rm BH}
=
\int_0^\infty
{\mathcal P}_<(\rh)\,\dd \rh
\ .
\label{Pbh}
\ee
\par
It appears natural to apply the horizon quantum mechanics to black hole geometries described
by coherent states and to verify under which conditions there exists a horizon with probability
close to one.
For this purpose, we will first reconstruct the state $\ket{\psi_{\rm S}}$ from the
effective energy density associated with the quantum corrected geometry~\eqref{Vq} in
Section~\ref{S:coherent};
using that result, we will obtain the horizon wavefunction in Section~\ref{S:hwf};
final remarks are given in Section~\ref{S:conc}.
\section{Coherent quantum states for Schwarzschild geometry}
\label{S:coherent}
\setcounter{equation}{0}
A metric of the form in Eq.~\eqref{metric} can be conveniently described as the mean field of the
coherent state of a (canonically normalised) free massless scalar field
$\sqrt{\gn}\,\Phi=(f-1)/2=V$ (see Ref.~\cite{Casadio:2021eio} for all the details).
From the Klein-Gordon equation
\begin{equation}
\left[
-\frac{\partial^2}{\partial t^2}
+
\frac{1}{r^2}\,\frac{\partial}{\partial r}
\left(r^2\,\frac{\partial}{\partial r}\right)
\right]
\Phi(t,r)
=
0
\ ,
\end{equation}
we obtain the (positive frequency) eigenfunctions
\begin{equation}\label{umodes}
 u_{k}
 =
 {e^{-i \, k \, t}}\,j_0 (k \, r)
\ ,
\end{equation}
where $j_{0}= \sin (k\,r) / k \, r$ with $k>0$ are spherical Bessel functions, which allow us
to write the field operator as
\be
\hat\Phi
=
\int\limits_0^{\infty}
\frac{
k^2
\,\dd k}{2\, \pi^2}
\,\sqrt{\frac {\hbar} {2 \, k}}
\left[
u_{k}\,
\hat{a}(k)
+
u_{k }^{\ast}\,
\hat{a}^{\dagger}(k)
\right]
\label{the quantum field operator}
\ee
and its conjugate momentum as
\be
\hat\Pi
=
i\int\limits_0^{\infty}
\frac{
k^2
\,\dd k}{2\, \pi^2}
\,\sqrt{\frac {\hbar \, k} {2 }}
\left[
u_{k}\,
\hat{a}(k)
-
u_{k }^{\ast}\,
\hat{a}^{\dagger}(k)
\right]
\
\label{the quantum field operator conjugate momentum}
\ee
where $\a$ and $\ac$ are the usual annihilation and creation operators.
\par
In particular, we are interested in a coherent state
\be
\ket{V_M}
=
e^{-N_M/2}\,
\exp\!\left\{
\int_0^\infty
\frac{k^2\,\dd k}{2\,\pi^2}\,g_k\,\ac(k)
\right\}
\ket{0}
\ ,
\ee
which effectively reproduces (as closely as possible) the Schwarzschild geometry, that is
\be
\label{the classical potential}
\sqrt{\gn}
\bra{V_M}
\hat\Phi(t, r )
\ket{V_M}
\simeq
V_M(r)
=
-\frac{2\,\gn\,M}{r}
\ .
\ee
From
\be
\label{spectral decomposition}
\bra{V_M}
\hat\Phi
\ket{V_M}
&\!\!=\!\!&
\int\limits_0^{\infty}
\frac {k^2\, \dd k} {2\,\pi^2}\,
\sqrt{\frac {2\,\lp\,\mpl}{k}}\,
g_{k} \, \cos(k\,t-\gamma_k)\,
j_0(k \,r)\,
\ ,
\ee
we impose $\gamma_k= k\, t$ for staticity and the coefficients $g_{k}$
can be determined by expanding the metric function $V_M=V_M(r)$
on the spatial part of the normal modes~\eqref{umodes}, to obtain
\be
g_{k}
=
-\frac{4\,\pi\,M}{\sqrt{2\,k^3}\,\mpl}
\ .
\label{value of V}
\ee
However, the corresponding normalisation factor
\be
\label{occupation number}
N_M
=
4\,\frac{M^2}{\mpl^2}
\int\limits_0^{\infty}
\frac{\dd k}{k}
\ee
diverges both in the infrared and in the ultraviolet.
The infrared divergence can be eliminated by embedding the geometry in a universe
of finite Hubble radius $r=R_\infty$, whereas the ultraviolet divergence could be removed by
assuming the existence of a matter core of finite size $r=R_{\rm s}$.
\par
For the present work, it is convenient to regularise the ultraviolet divergence
by replacing the coefficients in Eq.~\eqref{value of V} with
\be
g_{k}
=
-\frac {4\, \pi \, M\,e^{-\frac{k^2 \, R_{\rm s}^2} {4}}}{\sqrt{2 \, k^3}\,\mpl}
\ ,
\label{g_00}
\ee
which yields the total occupation number
\be
N_M
&\!\!=\!\!&
4\,\frac{M^2}{\mpl^2}
\int\limits_{R_\infty^{-1}}^{\infty}
\frac{\dd k}{k} \,
e^{- \frac {k^2 \, R_{s}^2} {2}}
\nonumber
\\
&\!\!=\!\!&
2 \,\frac{M^2}{\mpl^2}\,
\Gamma\!\left(0,\frac{R_{\rm s}^2}{2 \, R_{\infty}^2}
\right)
\nonumber
\\
&\!\!\simeq\!\!&
4 \,\frac{M^2}{\mpl^2}
\ln\!\left(\frac{R_{\infty}}{R_{\rm s}}
\right)
\ ,
\label{Nm}
\ee
where $\Gamma=\Gamma(a,x)$ is the incomplete gamma function and we assumed $R_{\rm s}\ll R_\infty$.
The coherent state $\ket{V_M}$ so defined corresponds to a quantum-corrected metric function
\be
V_{{\rm q}M}
=
\sqrt{\gn}
\bra{V_M}\hat\Phi\ket{V_M}
=
-\frac{\gn\,M}{r}
\, {\rm erf} \!\left(\frac{r}{R_{\rm s}}
\right)
\ ,
\label{VqM}
\ee
where $ {\rm erf} $ denotes the error function and we let $R^{-1}_{\infty} \to 0 $.
\subsection{Effective energy density}
From the definition of the mass function in Eq.~\eqref{M} and
\be
1+2\,V_{{\rm q}M}
=
1-\frac{2\,\gn\,m}{r}
\ ,
\ee
we easily obtain
\be
\rho(r)
=
-\frac{V_{{\rm q}M}}{4\,\pi\,\gn\,r^2}
\left(1+r\,\frac{V_{{\rm q}M}'}{V_{{\rm q}M}}\right)
\ .
\ee
\par
We next note that the quantum corrected potential~\eqref{VqM} is of the form
\be
V_{{\rm q}M}
=
V_M(r)\,v(r)
\ ,
\ee
where the function $v$ has the asymptotic behaviours
\be
v(r\to 0)\to 0
\qquad
{\rm and}
\qquad
v(r\gg R_{\rm s})\to 1
\ .
\label{Fasy}
\ee
The effective energy density therefore reads
\be
\rho
=
\frac{M\,v'}{4\,\pi\,r^2}
\ ,
\ee
so that Eq.~\eqref{Fasy} implies
\be
m(r\to\infty)
=
M
\int_0^\infty
v'(x)\,\dd x
=
M
\ ,
\ee
as expected.
\par
In particular, we have $v={\rm erf}({r}/{R_{\rm s}})$ and
\be
\label{effective energy density}
\rho
=
\frac {M\, e^{-\frac {r ^2 } { R_{\rm s}^2}}} { {2} \, \pi^{\frac {3} {2}} \, R_{\rm s}  \, r^2}
\ ,
\ee
which is the same result one would obtain from the Einstein field equations
$G^\mu_{\ \nu}=8\,\pi\,\gn\,T^\mu_{\ \nu}$, where $G^\mu_{\ \nu}$ is the Einstein
tensor for the quantum corrected metric from Eq.~\eqref{VqM}.
\section{Horizon quantum mechanics}
\label{S:hwf}
\setcounter{equation}{0}
We are interested in a matter source with energy density~\eqref{effective energy density}
made of a very large number $N$ of particles.
For simplicity, we assume that all particles are identical and have a mass $\mu = M / N$.
\par
The (normalised) wavefunction of each particle in position space can be estimated as
\begin{equation}
\label{wave function in position space solution 2}
\psi_{{\rm S }}(r _{i} )
\propto
\rho^{1/2}
\propto
\frac {e^{-\frac {r _{i} ^2 } {2 \, R_{\rm s}^2}} }
{\sqrt{2} \, \pi^{\frac {3} {4}} \, R_{\rm s}^{\frac {1} {2} } \, r _{i}}
\ ,
\end{equation}
where $i=1,\ldots,N$.
In momentum space, we then have
\be
\label{wave function in momentum space solution 2}
\psi_{{\rm S}}(k_{i})
=
\frac {2 \, \pi^{\frac {3} {4}} \, R_{\rm s}^{\frac {1} {2} } } {k_{i}} \,
{\rm erfi} \!\left( \frac {   k_{i} \,  R_{\rm s}} {\sqrt{2}}  \right)
\, e^{- \frac { k_{i}^2 \, R_{\rm s}^2  } { {2} }  }
\ ,
\ee
where ${\rm erfi}$ is the imaginary error function.
Notice that the wavefunction~\eqref{wave function in momentum space solution 2}
peaks around $k = R_\infty^{-1}$, and the imaginary error function can be approximated
for $k_i\,R_{\rm s}\ll 1$ as
\begin{equation}
\label{ApErfi}
{\rm erfi}\!\left( \frac {   k_{i} \,  R_{\rm s}} {\sqrt{2}}  \right)
\simeq
\sqrt{\frac{2}{\pi}} \,   k_{i} \,  R_{\rm s}
\ .
\end{equation}
Each particle can therefore be assumed in a state described by~\footnote{Given the approximation~\eqref{ApErfi},
the expression~\eqref{state in momentum space} ``underestimates'' the exact wavefunction at large $k\sim R_{\rm s}^{-1}$.
The related error can be reduced by decreasing the value of $R_{\rm s}$ with respect to the (unknown) actual size of the core.}
\be
\label{state in momentum space}
\ket{\psi_{\rm s}^{(i)}}
\simeq
\mathcal{N}_k
\int\limits_{R_\infty^{-1}}^{\infty}
{\dd k_i}\,
e^{- \frac { k_{i}^2 \, R_{\rm s}^2  } { {2} }  }
\,\ket{k_i}
\ ,
\ee
where $\mathcal{N}_k$ is a suitable normalisation factor.
\par
The dynamics of each particle is determined by a Hamiltonian $H_i$ with spectrum
\begin{equation}
\hat H_i
\ket{E_i}
=
E_i
\ket{E_i}
\ ,
\label{Hi}
\end{equation}
where
\be
E_{i}^2 = \mu^2+ \hbar^2 \, k_{i}^2
\ .
\ee
Thus, we can rewrite the state~\eqref{state in momentum space} of each particle as
\be
\ket{\psi_{\rm s}^{(i)}}
&\!\!\simeq\!\!&
\mathcal{N}_E
\int\limits_{\mu}^{\infty}
\dd E_i \,
e^{- \frac {\left( E_i^2 - \mu^2 \right) \, R_{s}^2} {2 \,\mpl^2\, \lp^2}} \,
\ket{E_i}
\ ,
\ee
where $\mathcal{N}_E$ is also a normalisation factor.
\par
The total wavefunction of the source will be given by the symmetrised product of $N$ such states,
\be
\ket{\psi_{N}}
\simeq
\frac { 1 } {N!} \,
\sum_{ \{\sigma_i \}}^{N} \,
\left[
\bigotimes_{i=1}^{N}
\ket{\psi_{\rm s}^{(i)}}
\right]
\ ,
\label{total wave-function}
\ee
where the sum is over all the permutations $\{\sigma_i \}$ of the $N$ states.
\subsection{Source spectral decomposition}
\label{SS:CE}
The above $\ket{\psi_{N}}$ can be decomposed into eigenstates $\ket{E}$ of the
total Hamiltonian~\footnote{Notice that we are assuming that the total energy is just the
sum of individual particle energies to parallel the expression~\eqref{M} of the classical
mass function.}
\be
H
=
\sum_{i=1}^N H_i
=
\sum_{i=1}^N
\left(
\mu^2+ \hbar^2 \, k_{i}^2
\right)^{1/2}
\ .
\ee
The details of the (approximate analytical) calculation are shown in Appendix~\ref{A:spectrum}, where
we find that $C(E)\equiv\langle E\ket{\psi_{N}}\simeq 0$, for $E<M$, and
\be
C(E)
\simeq
\mathcal{N}_c
\left(\frac{E-M}{\mpl}\right)^{M/\mu}
e^{-\frac {R_{\rm s}^2\,\mu\,(E-M)} {\lp^2\,\mpl^2}}
\ ,
\ee
for $E>M$, with the normalisation constant $\mathcal{N}_c=\mathcal{N}_+$ given
in Eq.~\eqref{N+}.
This result means that we can describe the quantum state $\ket{\psi_{N}}$
of our $N$-particle system by means of the effective one-particle state
\be
\ket{\Psi_{\rm S}}
\simeq
\mathcal{N}_{\rm S}
\int\limits_{M}^{\infty}
\dd E  \,
\left(\frac{E-M}{\mpl}\right)^{M/\mu}
e^{-\frac {R_{\rm s}^2\,\mu\,(E-M)} {\lp^2\,\mpl^2}}\,\ket{E}
\ ,
\label{single-particle state}
\ee
with $E^2 = M^2 + \hbar^2\, k^2$ and $\mathcal{N}_{\rm S}$ is a normalisation
constant.
For example, the expectation value of the total energy can be approximated 
with  its upper bound computed in Eq.~\eqref{A<H>} and reads
\be
\expec{\hat H}
\simeq
M
\left(
1
+
\frac {\mpl^2\,\lp^2} {\mu^2\,R_{\rm s}^2}
\right)
=
M
\left(
1
+
\frac {\lambda_\mu^2}{R_{\rm s}^2}
\right)
\ ,
\label{<H>}
\ee
where $\lambda_\mu$ is the Compton length of the constituent particles of mass $\mu$.
We notice that the relative correction becomes negligibly small for $R_{\rm s}\gg \lambda_\mu$
and diverges for $R_{\rm s}\to 0$.
This is another indication that no well-defined coherent state exists for a pure Schwarzschild
geometry~\cite{Casadio:2021eio}.
\subsection{Horizon wavefunction}
\label{SS:hwf}
We can now obtain the horizon wavefunction from the effective single-particle
wavefunction~\eqref{single-particle state} by setting $r_{\rm H} = 2\,\gn\, E$
and defining $\ket{\rh}\propto\ket{2\,\ell_{\rm p}\, E / m_{\rm p}}$.
This yields $\Psi_{\rm H}(\rh)\simeq 0$, for $\rh<\Rh=2\,\gn\,M$, and
\be
\Psi_{\rm H}(\rh)
\simeq
\mathcal{N}_{\rm H}
\left(\frac{\rh-\Rh}{\lp}\right)^{\frac{\mpl\,\Rh}{2\,\mu\,\lp}}
e^{-\frac {\mu\left( r_{\rm H} - R_{\rm H} \right)R_{\rm s}^2} {2\,\mpl\,\lp^3}}
\ ,
\label{psi_H}
\ee
for $\rh\ge \Rh$, where the normalisation constant $\mathcal{N}_{\rm H}$
is given in Eq.~\eqref{Nh}.
\par
The expectation value of the gravitational radius is computed in Eq.~\eqref{<rh>} and
can be written as
\be
\expec{\hat r_{\rm H}}
\simeq
\Rh
\left(
1
+
\frac {\lambda_\mu^2}{R_{\rm s}^2}
\right)
\ ,
\label{exprh}
\ee
which is in perfect agreement with the expression of the energy given in Eq.~\eqref{<H>}.
It is again noteworthy that $\expec{\hat r_{\rm H}}>\Rh$, although the correction with respect to the
classical expression is negligible for an astrophysical black hole unless the core is of a size
comparable to the Compton length $\lambda_\mu$.
It is also important to recall that $\expec{\hat r_{\rm H}}$ is the horizon radius
only if the core is sufficiently smaller, as we will determine next.
\par
\begin{figure}[t]
\centering
\includegraphics[width=9cm]{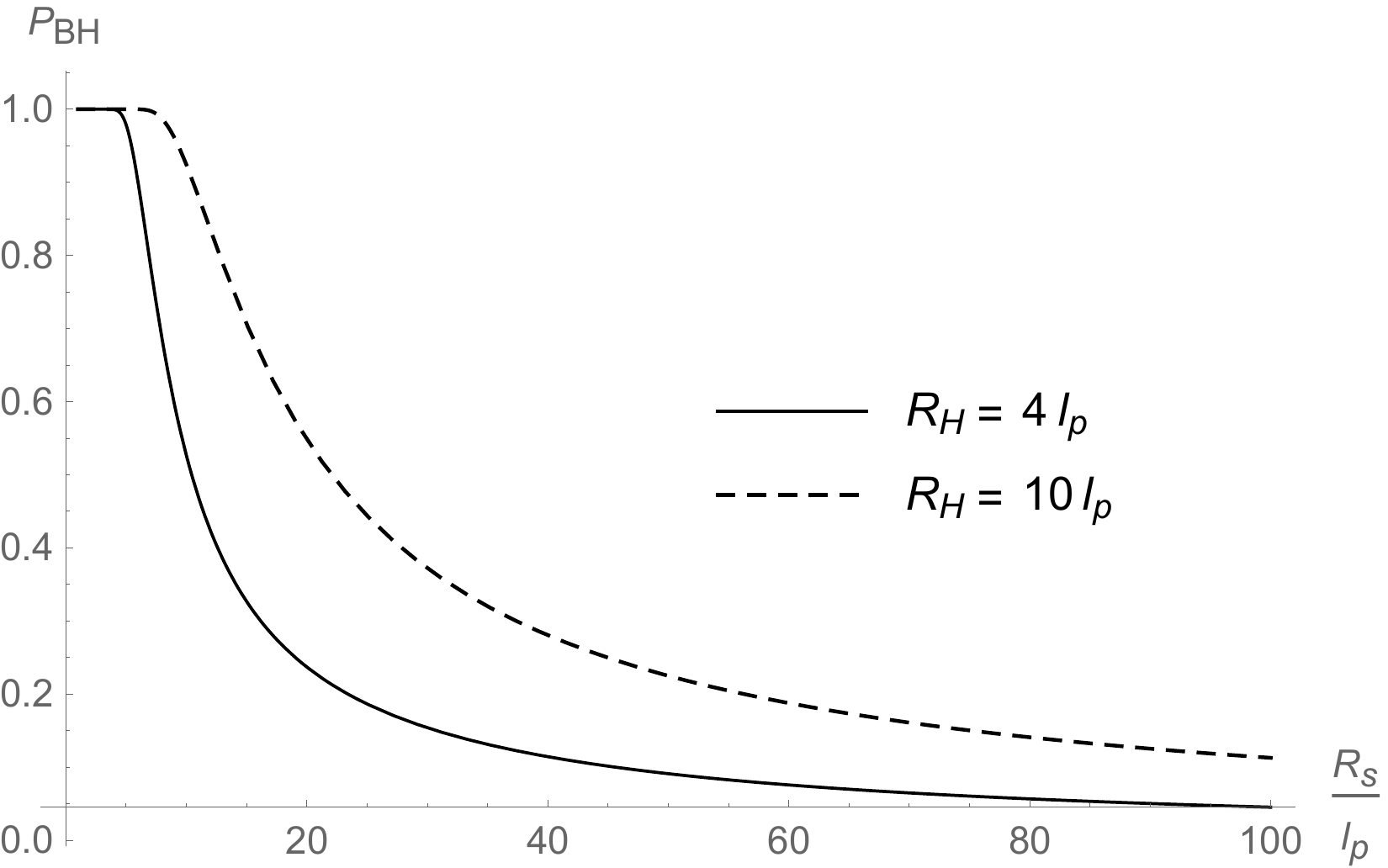}
\caption{Probability that the coherent state is a black hole as a function of $R_{\rm s}$
for different values of $\Rh$ (and same value of $\mu=0.2\,\mpl$).}
\label{f:P1}
\end{figure}
\begin{figure}[t]
\centering
\includegraphics[width=9cm]{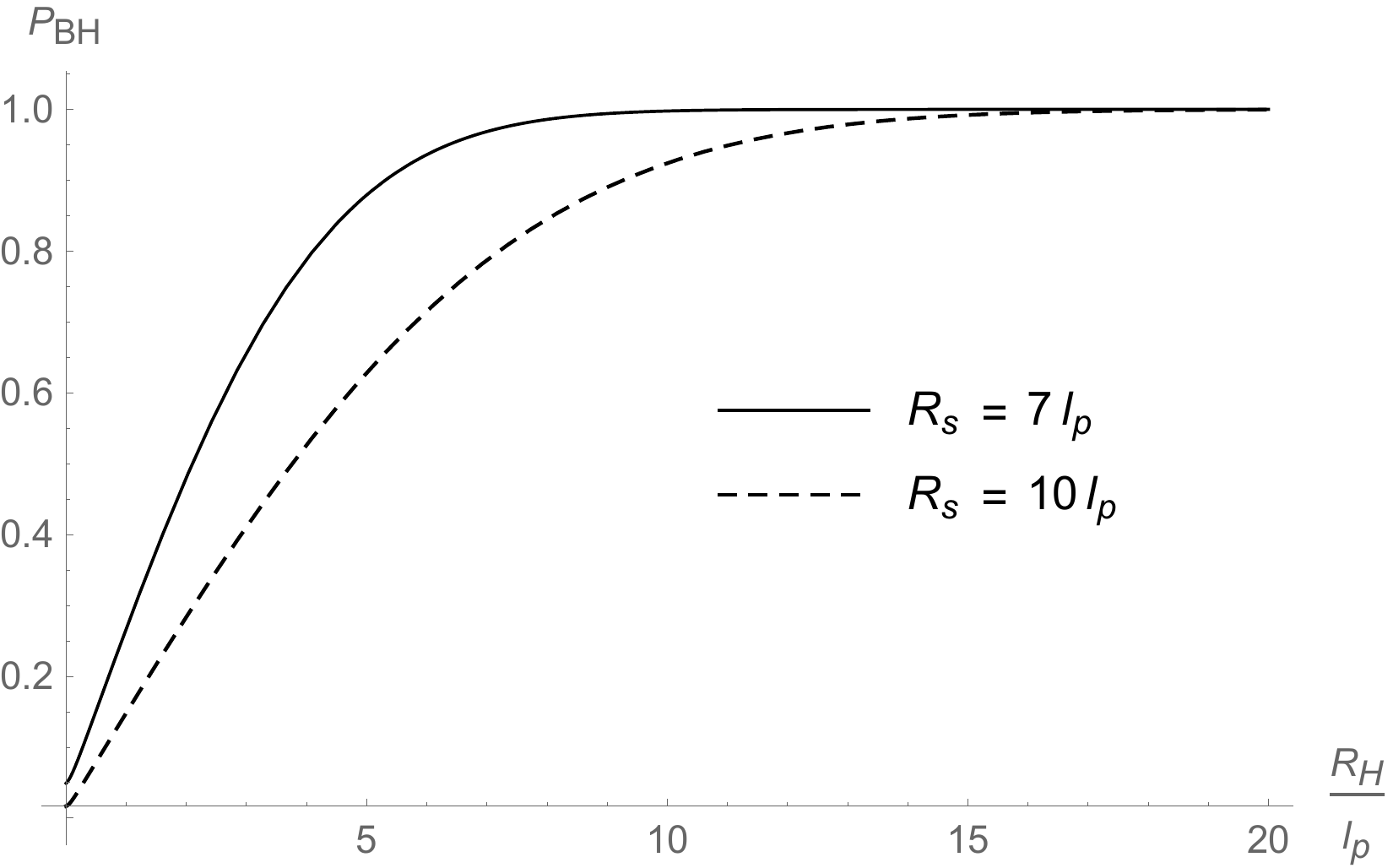}
\caption{Probability that the coherent state is a black hole as a function of $\Rh$ for
different values of ${ R_{\rm s}}$ (and same value of $\mu=0.2\,\mpl$).}
\label{f:P2}
\end{figure}
\begin{figure}[t]
\centering
\includegraphics[width=9cm]{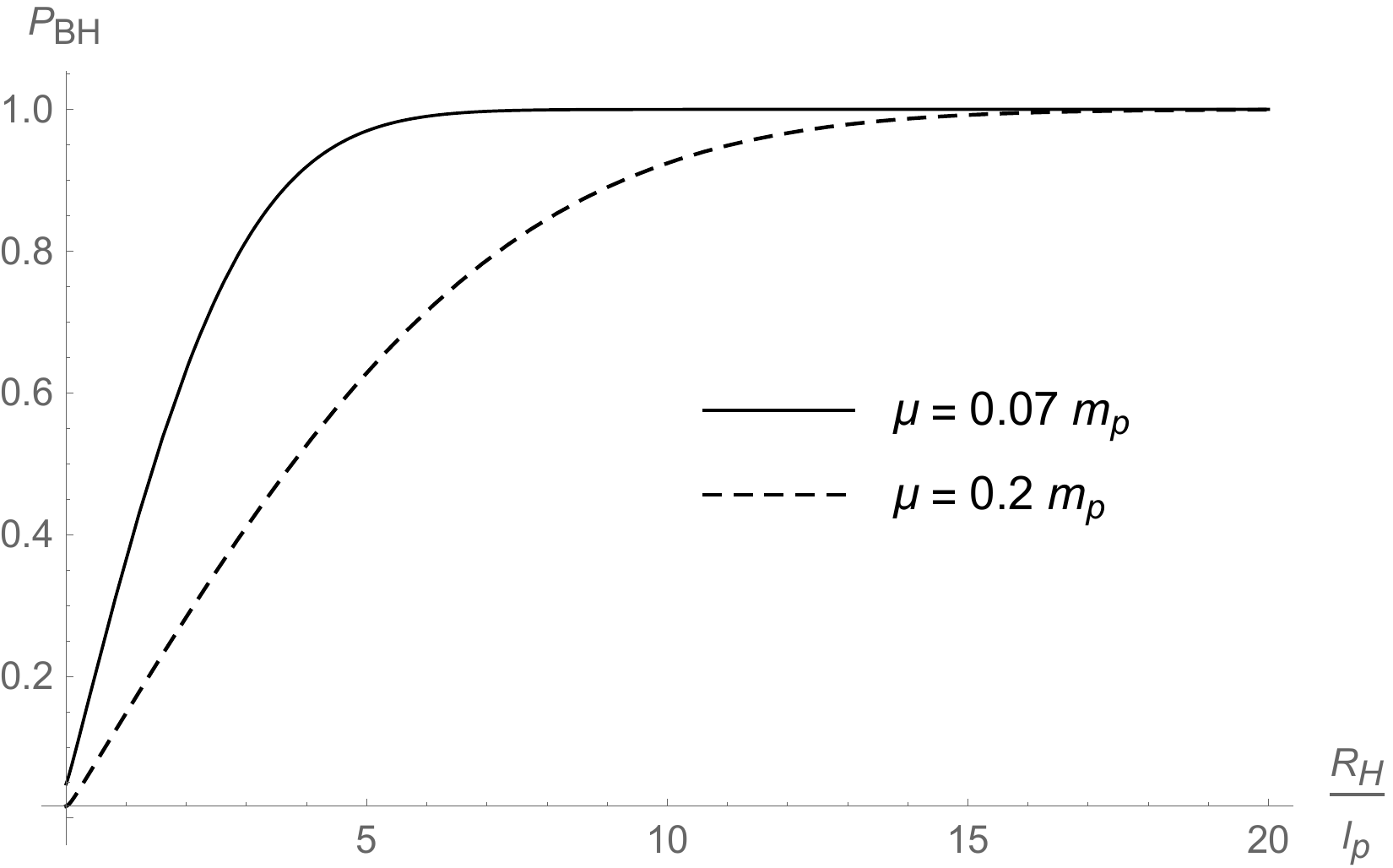}
\caption{Probability that the coherent state is a black hole as a function of $\Rh$ for
different values of ${ \mu}$ (and same value of $R_{\rm s}=10\,\lp$).}
\label{f:P3}
\end{figure}

By means of the effective single-particle wavefunction~\eqref{single-particle state} and the
horizon wavefunction~\eqref{psi_H}, we can numerically compute the probability $P_{\rm BH}$
defined in Eq.~\eqref{Pbh} that the system lies inside its own gravitational radius and is a black hole,
as reviewed in Section~\ref{S:intro}.
More details of the calculation are given in Appendix~\ref{A:norm}, where we show that the final
expression of $P_{\rm BH}$ can only be estimated numerically.
Some cases are displayed in Figs.~\ref{f:P1}-\ref{f:P3}, with values of $\Rh$, $R_{\rm s}$
and $\mu$ chosen for clarity, albeit they fall far from any astrophysical regimes.
From those graphs, it appears that the probability increases for decreasing size $R_{\rm s}$
of the core and for increasing (decreasing) mass $M$ ($\mu$)
(equivalent to increasing $\Rh=2\,\gn\,M$ or the number $N=M/\mu$ of matter particles).
For example, a core of size $R_{\rm s}=10\,\lp$ can be a black hole of radius $\Rh=10\,\lp$
with probability $P_{\rm BH}\gtrsim 0.9$ but this probability drops to $P_{\rm BH}\lesssim 0.5$
if $\Rh=4\,\lp$.
This result is in qualitative agreement with the expectation~\eqref{exprh} for very massive
black holes with cores larger than $\lambda_\mu$, but smaller than the classical gravitational
radius $\Rh$.
\section{Conclusions and outlook}
\label{S:conc}
\setcounter{equation}{0}
We have here employed the formalism of the horizon quantum mechanics~\cite{Casadio:2013tma}
in order to verify that coherent state black hole geometries of the Schwarzschild type sourced by a core
of large mass $M$ and with a size $R_{\rm s}$ larger than Planckian are very likely to display
an outer horizon and be black holes in the usual sense.
For that purpose, we needed to find an explicit description of the (electrically neutral and
spherically symmetric) matter core in terms of a many-particle state that was then expressed
as a superposition of total energy eigenstates.
\par
Our analysis supports the conclusion that the system is indeed a black hole of mass $M\gg\mpl$
if its core made of particles of mass $\mu$ has a size
$R_{\rm s}\gtrsim \lambda_\mu\gg\lp$ but (sufficiently)
smaller than the classical gravitational radius $\Rh=2\,\gn\,M$.
It would be interesting to generalise the above analysis to include electric charge and
rotation.
Whereas the former case should be rather straightforward, including rotation is going to be much more
problematic since it will require extending the horizon quantum mechanics beyond the perturbative
regime considered in Ref.~\cite{Casadio:2017nfg}.
\par
Clearly, the emerging picture is that the (location of the) horizon in quantum physics
is fuzzy and, at least for sufficiently large matter cores, one would have ``quasi'' black hole geometries.
In such a picture, the late stage of binary black hole mergers would open a window
into quantum features of the gravitational collapse that might affect the emission of gravitational
waves at the very peak, or during the ring-down phase, for instance by affecting the black hole
Love numbers~\cite{Cardoso:2017cfl} and possible echos~\cite{Cardoso:2017cqb,Buoninfante:2020tfb}.
Of course, quantitative predictions for such effects would require specific analysis that go beyond
the scope of the present work.
\section*{Acknowledgments}
W.F.~acknowledges the financial support provided by the scholarship granted by the
Chinese Scholarship Council (CSC).
A.G.~is supported in part by the Science and Technology Facilities Council
(grants n.~ST/T006048/1 and ST/Y004418/1).
W.F.~and R.C.~are partially supported by the INFN grant FLAG.
The work of A.G.~and R.C.~has also been carried out in the framework of activities of the
National Group of Mathematical Physics (GNFM, INdAM).
\section*{Data Availability}
No datasets were generated or analysed during the current study.
\section*{Conflict of interest}
The authors declare that they have no conflict of interest.
\appendix
\section{Spectral decomposition and total energy}
\label{A:spectrum}
\setcounter{equation}{0}
Here we show how the total wavefunction~\eqref{total wave-function}, that is
\be
\ket{\psi_{N}}
\simeq
\frac { 1 } {N!} \,
\sum_{ \{\sigma_i \}}^{N} \,
\left[
\bigotimes_{i=1}^{N}
\mathcal{N}_E
\int\limits_{\mu}^{\infty}
\dd E_i \,
e^{- \frac {\left( E_i^2 - \mu^2 \right) \, R_{s}^2} {2 \,\mpl^2\, \lp^2}} \,
\ket{E_i}
\right]
\ ,
\label{psiN}
\ee
can be decomposed in terms of the total energy eigenstates $\ket{E}$ by
computing the spectral coefficients $C(E) \equiv \pro{E}{\psi_{N}}$.
From Eq.~\eqref{psiN}, we first find
\be
C(E)
&\!\!=\!\!&
\frac { 1 } {N!} \,
\bra{E}
\sum_{ \{\sigma_i \}}^{N} \,
\left[
\bigotimes_{i=1}^{N}
\mathcal{N}_E
\int\limits_{\mu}^{\infty}
\dd E_i \,
e^{- \frac {\left( E_i^2 - \mu^2 \right) \, R_{s}^2} {2 \,\mpl^2\, \lp^2}} \,
\ket{E_i}
\right]
\nonumber
\\
&\!\!=\!\!&
\frac { \mathcal{N}_E^{N} } {N!}
\int\limits_{\mu}^{\infty}   \dd E_{1}
\cdots
\int\limits_{\mu}^{\infty}   \dd E_{N}
\left[\prod_{i=1}^{N} \, {
e^{- \frac {\left( E_i^2 - \mu^2 \right) \, R_{s}^2} {2 \,\mpl^2\, \lp^2}} }
\right]
\delta\!
\left(
E-\sum_{i=1}^{N}E_{i}
\right)
\ .
\ee
Since $\sum_{i=1}^N E_i\ge N\,\mu=M$, it follows that $C(E<M)=0$.
For $E\ge M$, we can use $E_N=\sum_{i=1}^{N-1} E_i$ and write
\be
C(E)
\propto
\int\limits_{\mu}^{\infty}   \dd E_{1}\cdots
\int\limits_{\mu}^{\infty}   \dd E_{N-1} \,
{\rm exp}\left\{
- \sum_{i=1}^{N-1} \frac {\left( E_i^2 - \mu^2 \right) R_{s}^2} {2 \,\mpl^2\, \lp^2}
-
\frac { \left[ \left(  E - \sum\limits_{i=1}^{N-1} \, E_{i}  \right)^2 -\mu^2  \right]
R_{s}^2} {2 \,\mpl^2\, \lp^2}
\right\}
\ .
\qquad
\label{CE1}
\ee
It is now convenient to define the function
\be
F(E,{E_{i}})
&\!\!\equiv\!\!&
\sum_{i=1}^{N-1}  {\left( E_i^2 - \mu^2 \right)  }
+
\left(  E - \sum_{i=1}^{N-1} \, E_{i}  \right)^2 - \mu^2
\nonumber
\\
&\!\!=\!\!&
\sum_{i=1}^{N-1}
{\left( \mathcal{E}_i + \mu \right)^2 -  N \,\mu^2}
+
\left[  E - \sum_{i=1}^{N-1} \, \mathcal{E}_{i} -  (N-1)\, \mu\right]^2
\ ,
\ee
where $\mathcal{E}_i = E_{i} -\mu$.
By recalling that $M = N \, \mu$, we then obtain
\be
F(E,E_i)
&\!\!=\!\!&
\sum_{i=1}^{N-1}
\left( \mathcal{E}_i + \mu \right)^2 -  \mu\,M
+
\left[  \left(E-M\right) - \sum_{i=1}^{N-1} \, \mathcal{E}_{i} +\mu\right]^2
\nonumber
\\
&\!\!=\!\!&
\left(  E - M \right)^2
+
\left( \sum_{i=1}^{N-1} \, \mathcal{E}_{i}  \right)^2
+
\mu^2
- 2 \left(  E - M \right) \sum_{i=1}^{N-1} \, \mathcal{ E}_{i}
+
2 \,\mu\left(  E - M \right)
-
2\,\mu \,\sum_{i=1}^{N-1} \, \mathcal{ E}_{i}
\nonumber
\\
&&
+
\sum_{i=1}^{N-1}
\mathcal{E}_{i}^2
+2 \, \mu \, \sum_{i=1}^{N-1}\mathcal{E}_{i}
+
(N-1)\,\mu^2
-
\mu\,M
\nonumber
\\
&\!\!=\!\!&
\left(  E - M \right)^2
+
2\,\mu\left(  E - M \right)
-
2 \left(  E - M \right) \sum_{i=1}^{N-1}  \mathcal{E}_{i}
+
\left( \sum_{i=1}^{N-1} \, \mathcal{E}_{i}  \right)^2
+
\sum_{i=1}^{N-1}\mathcal{E}_i^2
\nonumber
\\
&\!\!=\!\!&
\left[E-(M-\mu)\right]^2
-
\mu^2
-
2 \left(  E - M \right) \sum_{i=1}^{N-1}  \mathcal{E}_{i}
+
\left( \sum_{i=1}^{N-1} \, \mathcal{E}_{i}  \right)^2
+
\sum_{i=1}^{N-1}\mathcal{E}_i^2
\ .
\ee
Plugging this result into Eq.~\eqref{CE1} yields
\be\label{CEIEM}
C(E)
&\!\!\propto\!\!&
e^{-\frac {R_{\rm s}^2} {2 \,\lp^2\,\mpl^2}
\left\{\left[E-(M-\mu)\right]^2
-
\mu^2\right\}
}
\int\limits_{0}^{\infty}   \dd \mathcal{E}_{1}\cdots
\int\limits_{0}^{\infty}   \dd \mathcal{E}_{N-1}
\nonumber
\\
&&
\times
\exp \left\{
\frac {R_{s}^2} {2\,\lp^2\,\mpl^2}
\left[
2\left(  E - M \right) \sum_{i=1}^{N-1} \, \mathcal{E}_i
-
\left( \sum_{i=1}^{N-1} \, \mathcal{E}_i  \right)^2
-
\sum_{i=1}^{N-1}\mathcal{E}_i^2
\right]
\right\}
\nonumber
\\
&\!\!\equiv\!\!&
e^{-\frac {R_{\rm s}^2} {2 \,\lp^2\,\mpl^2}
\left\{\left[E-(M-\mu)\right]^2
-
\mu^2\right\}
}\,
I(E, M)
\ .
\ee
We next note that, since $\mathcal{E}_i\ge0$ for $i=1,\ldots,N-1$, we have
\be
0
\le
\sum_{i=1}^{N-1}\mathcal{E}_i^2
\le
\left( \sum_{i=1}^{N-1} \, \mathcal{E}_i  \right)^2
\ .
\label{0EE2}
\ee
A lower bound $I_-\le I$ is obtained from the upper bound $\sum_i\mathcal E_i^2=(\sum_i\mathcal E_i)^2$
in Eq.~\eqref{0EE2} and is given by
\be
I_-
&\!\!=\!\!&
\int\limits_{0}^{\infty}   \dd \mathcal{E}_{1}\cdots
\int\limits_{0}^{\infty}   \dd \mathcal{E}_{N-1}
\exp\left[
\frac{R_{s}^2\,(E-M)}{\lp^2\,\mpl^2}\,\sum_{i=1}^{N-1}\mathcal{E}_i
\right]
\exp \left[
-\frac {R_{s}^2} {\lp^2\,\mpl^2}
\left(\sum_{i=1}^{N-1}\mathcal{E}_i\right)^2
\right]
\nonumber
\\
&\!\!\propto\!\!&
\int\limits_{0}^{\infty}
\mathcal{E}^{N-2}
\dd \mathcal{E}
\exp \left[
-\frac {R_{s}^2\,\mathcal{E}^2} {\lp^2\,\mpl^2}
+
\frac{R_{s}^2\,(E-M)\,\mathcal{E}}{\lp^2\,\mpl^2}
\right]
\nonumber
\\
&\!\!=\!\!&
\frac {1} {2} \left( \frac {R_{\rm s}} {\mpl\,\lp} \right)^{1-N}
\Gamma\!\left(\frac {N-1} {2}\right)
\,_{1}F_{1}\!\left[ \frac {N-1} {2}, \frac {1} {2}, \frac {\left(  E-M \right)^2 \, R_{\rm s}^2} { 4 \, \mpl^2\,\lp^2} \right]
\nonumber
\\
&&
+
\frac {1} {2} \left( \frac {R_{\rm s}} {\mpl\,\lp} \right)^{1-N}
\frac {\left(  E-M \right)  \, R_{\rm s} } {  \mpl \,\lp }\,
\Gamma\!\left(  \frac {N} {2} \right)
\,_{1}F_{1}\!\left[  \frac {N} {2}  , \frac {3} {2} , \frac {\left(  E-M \right)^2 \, R_{\rm s}^2} { 4 \, \mpl^2\,\lp^2} \right]
\ ,
\label{DoB}
\ee
where $\Gamma$ is the Euler gamma function and $ _{1}F_{1} $ the Kummer confluent hypergeometric function.
An upper bound $I\le I_+$ is likewise obtained from the lower bound $\sum_{i}\mathcal E_i^2=0$ in
Eq.~\eqref{0EE2} and is given by
\be
I_+
&\!\!=\!\!&
\int\limits_{0}^{\infty}   \dd \mathcal{E}_{1}\cdots
\int\limits_{0}^{\infty}   \dd \mathcal{E}_{N-1}
\exp\left[
\frac{R_{s}^2\,(E-M)}{\lp^2\,\mpl^2}\,\sum_{i=1}^{N-1}\mathcal{E}_i
\right]
\exp \left[
-\frac {R_{s}^2} {2\,\lp^2\,\mpl^2}
\left(\sum_{i=1}^{N-1}\mathcal{E}_i\right)^2
\right]
\nonumber
\\
&\!\!\propto\!\!&
\int\limits_{0}^{\infty}
\mathcal{E}^{N-2}
\dd \mathcal{E}
\exp \left[
-\frac {R_{s}^2\,\mathcal{E}^2} {2\,\lp^2\,\mpl^2}
+
\frac{R_{s}^2\,(E-M)\,\mathcal{E}}{\lp^2\,\mpl^2}
\right]
\nonumber
\\
&\!\!=\!\!&
\frac {1} {2^{\frac {N} {2}- \frac {3} {2}}}
\left( \frac {R_{\rm s}} {\mpl\,\lp} \right)^{1-N}
\Gamma\!\left(\frac {N-1} {2}\right)
\,_{1}F_{1}\!\left[ \frac {N-1} {2}, \frac {1} {2} , \frac {\left(  E-M \right)^2 \, R_{\rm s}^2} { 2 \, \mpl^2\,\lp^2} \right]
\nonumber
\\
&&
+
\frac {1} {2^{\frac {N} {2}- 1}}
\left( \frac {R_{\rm s}} {\mpl\,\lp} \right)^{1-N}
\frac {\left(  E-M \right)  \, R_{\rm s} } {  \mpl \,\lp }\,
\Gamma\!\left(  \frac {N} {2} \right)
\,_{1}F_{1}\!\left[  \frac {N} {2}  , \frac {3} {2} , \frac {\left(  E-M \right)^2 \, R_{\rm s}^2} { 2 \, \mpl^2\,\lp^2} \right]
\ .
\label{UpB}
\ee
\par
For $R_{\rm s}\gg\lp$ and $(E-M)\gtrsim\mpl$, we can employ the asymptotic behaviour of the Kummer
confluent hypergeometric function,
\be
_{1}F_{1}\,\left( a  , b , x\right)
\sim
x^{a-b}\,e^{x}
\ ,
\label{1F1inf}
\ee
for $x\sim (E-M)^2\,R_{\rm s}^2/\lp^2\,\mpl^2\gg 1$, which leads to
\be
I_-
&\!\!\simeq\!\!&
\frac {1} {2^{N - 2}}\,
\frac {\mpl\,\lp} {R_{\rm s}}
\left(E-M\right)^{N-2}\,
e^{\frac {\left(  E-M \right)^2 R_{\rm s}^2} { 4 \, \mpl^2\,\lp^2} }
\left[
2\,\Gamma\!\left(\frac {N-1} {2}\right)
+
\Gamma\!\left(\frac {N} {2}\right)
\right]
\nonumber
\\
&\!\!\simeq\!\!&
\Gamma\!\left(\frac {N-1} {2}\right)
\frac {\mpl\,\lp} {2^{N - 1}\,R_{\rm s}}
\left(E-M\right)^{N-2}
e^{\frac {\left(  E-M \right)^2 R_{\rm s}^2} { 4 \, \mpl^2\,\lp^2} }
\ ,
\ee
where we also used $2\,\Gamma \left((N-1)/2\right)>\Gamma\left(N/2\right)$
for $N\gg 1$ in the last step.
Likewise,
\be
I_+
&\!\!\simeq\!\!&
\frac {\mpl\,\lp} {\sqrt{2}\,R_{\rm s}} \,
\left(E-M\right)^{N-2}\,
e^{\frac {\left(  E-M \right)^2 R_{\rm s}^2} { 2 \, \mpl^2\,\lp^2} }
\left[
\Gamma \left(\frac {N-1} {2}\right)
+
2\,\Gamma \left(\frac {N} {2}\right)
\right]
\nonumber
\\
&\!\!\simeq\!\!&
\sqrt{2}\,
\Gamma \left(\frac {N} {2}\right)
\frac {\mpl\,\lp} {R_{\rm s}} \,
\left(E-M\right)^{N-2}\,
e^{\frac {\left(  E-M \right)^2 R_{\rm s}^2} { 2 \, \mpl^2\,\lp^2} }
\ ,
\ee
where we used $2\,\Gamma\!\left(N/2\right)>\Gamma\!\left((N-1)/2\right)$
for $N\gg 1$.
\par
Therefore, we have
\be
\Gamma\! \left(\frac {N-1} {2}\right)
\frac {\mpl\,\lp} {2^{N - 1}\,R_{\rm s}}
\left(E-M\right)^{N-2}
e^{\frac {\left(  E-M \right)^2 R_{\rm s}^2} { 4 \, \mpl^2\,\lp^2} }
\lesssim
I
\lesssim
\sqrt{2}\,
\Gamma\! \left(\frac {N} {2}\right)
\frac {\mpl\,\lp} {R_{\rm s}}
\left(E-M\right)^{N-2}
e^{\frac {\left(  E-M \right)^2 R_{\rm s}^2} { 2 \, \mpl^2\,\lp^2} }
\ .
\ee
We might note that the above approximation fails for $0<(E-M)\lesssim\mpl$ at fixed value
of $R_{\rm s}\gtrsim\lp$, for which we instead find the Taylor expansion
\be
I_+
\sim
I_-
\propto
1
+
\mathcal{O}\left(\frac{E-M}{\mpl}\right)
\ .
\ee
However, for an astrophysical system of mass $M\gg\mpl$, this regime can be discarded overall.
\par
By recalling that $N=M/\mu\gg 1$, we finally obtain the bounding functions
\be
C_-(E)
=
\mathcal{N}_-
\left(\frac{E-M}{\mpl}\right)^{M/\mu}
e^{-\frac {R_{\rm s}^2\,\mu\,(E-M)} {\lp^2\,\mpl^2}}\,
e^{-\frac {R_{\rm s}^2\left(  E-M \right)^2} { 4 \, \mpl^2\,\lp^2} }
\label{C-}
\ee
and
\be
C_+(E)
&\!\!=\!\!&
\mathcal{N}_+
\left(\frac{E-M}{\mpl}\right)^{M/\mu}
e^{-\frac {R_{\rm s}^2\,\mu\,(E-M)} {\lp^2\,\mpl^2}}
\ .
\label{C+}
\ee
The normalizations $\mathcal{N}_{\pm}$ can be obtained from the condition
\be
1
=
\int\limits_{M}^{\infty}
C^2_\pm(E)  \,\dd E
\ ,
\label{NNc}
\ee
yielding
\be
\mathcal{N}_{-}^{-2}
&\!\!=\!\!&
\left(\frac{\lp}{\sqrt{2}\,R_{\rm s}}\right)^{\frac{2\,M}{\mu}}
\Gamma\!\left(1+\frac{2\,M}{\mu}\right)
U\!\left(1+
\frac{M}{ \mu}, \frac{3}{2},\frac{2\,\mu^2\,R_{\rm s}^2}{\mpl^2\,\lp^2}\right)
\nonumber
\\
&\!\!\simeq\!\!&
\left(\frac{\lp}{\sqrt{2}\,R_{\rm s}}\right)^{\frac{2\,M}{\mu}}
\Gamma\!\left(\frac{2\,M}{\mu}\right)
U\!\left(\frac{M}{\mu},\frac{3}{2},\frac{2\,\mu^2\,R_{\rm s}^2}{\mpl^2\,\lp^2}\right)
\label{N-}
\ ,
\ee
where $U=U(a,b,x)$ is the Tricomi confluent hypergeometric function, and
\be
\mathcal{N}_{+}^{-2}
&\!\!=\!\!&
\mpl
\left(
\frac{\mpl\,\lp^2} {2\,\mu\,R_{\rm s}^2}
\right)^{1+\frac{2\,M}{\mu}}
\Gamma\!\left(1+\frac{2\,M}{\mu}\right)
\nonumber
\\
&\!\!\simeq\!\!&
\mpl
\left(
\frac{\mpl\,\lp^2}{2\,\mu\,R_{\rm s}^2}
\right)^{\frac{2\,M}{\mu}}
\Gamma\!\left(\frac{2\,M}{\mu}\right)
\ .
\label{N+}
\ee
\par
In Section~\ref{SS:CE}, we use the upper bounding function~\eqref{C+}
in order to estimate the maximum corrections to the total energy.
In fact, the bounds from the spectral coefficients can be used to bound
the expectation value of the total energy as
\be
M+H_-
\lesssim
\expec{\hat H}
\lesssim
M+H_+
\ ,
\ee
where
\be
H_+
&\!\!=\!\!&
\int\limits_{M}^{\infty}
C_+^2(E) \, E\,\dd E
-
M
\nonumber
\\
&\!\!=\!\!&
\int\limits_{0}^{\infty}
C_+^2(\mathcal{E}) \, \mathcal{E}\,\dd \mathcal{E}
\nonumber
\\
&\!\!=\!\!&
\frac{\lp^2\,\mpl^2\,(2\,M+\mu)\,}
{2\,\mu^2\,R_{\rm s}^2}
\nonumber
\\
&\!\!\simeq\!\!&
M\,
\frac{\mpl^2\,\lp^2}
{\mu^2\,R_{\rm s}^2}
\label{dHp}
\ee
and
\be
H_-
=
\int\limits_{0}^{\infty}
C_-^2(\mathcal{E}) \, \mathcal{E}\,\dd \mathcal{E}
=
\frac{
\lp^2\,\mpl^2\,(2\,M+\mu)\,
U\!\left(1+\frac{M}{\mu},\frac{1}{2},\frac{2\,\mu^2\,R_{\rm s}^2}{\mpl^2\,\lp^2}\right)}
{2\,\mu^2\,R_{\rm s}^2\,
U\!\left(1+\frac{M}{\mu},\frac{3}{2},\frac{2\,\mu^2\,R_{\rm s}^2}{\mpl^2\,\lp^2}\right)}
\ .
\label{dHm}
\ee
Employing the definition of the Tricomi confluent hypergeometric function,
\be
U(a,b,x)
=
\frac {\Gamma\left( 1-b \right)} {\Gamma\left(  a+1-b \right)} \,
_{1}F_{1}\,\left( a  , b , x\right)
+
\frac {\Gamma\left( b-1 \right)} {\Gamma\left(  a \right)} \,
x^{1-b} \,
_{1}F_{1}\,\left( a+1-b  , 2 - b , x\right)
\ ,
\ee
and the asymptotic behaviour of the Kummer confluent hypergeometric function~\eqref{1F1inf},
we find
\be
U\!\left(1+\frac{M}{\mu},\frac{1}{2},\frac{2\,\mu^2\,R_{\rm s}^2}{\mpl^2\,\lp^2}\right)
\simeq
\left[ \frac {1} {\Gamma\!\left(  \frac{M}{\mu}+ \frac {3} {2} \right)} -\frac {2} {\Gamma\!\left( \frac{M}{\mu}+1 \right)} \right]
\Gamma\!\left( \frac {1} {2} \right)
\left( \frac{\sqrt{2 }\,\mu\,R_{\rm s}}{\mpl\,\lp}  \right)^{1+2 \, \frac{M}{\mu}}
e^{\frac{2\,\mu^2\,R_{\rm s}^2}{\mpl^2\,\lp^2}}
\ee
and
\be
U\!\left(1+\frac{M}{\mu},\frac{3}{2},\frac{2\,\mu^2\,R_{\rm s}^2}{\mpl^2\,\lp^2}\right)
\simeq
\left[ \frac {1} {\Gamma\!\left(  \frac{M}{\mu}+ 1 \right)} -\frac {2} {\Gamma\!\left(  \frac{M}{\mu}+\frac {1} {2} \right)} \right]
\Gamma\left( \frac {1} {2} \right)
\left( \frac{\sqrt{2 }\,\mu\,R_{\rm s}}{\mpl\,\lp}  \right)^{2 \, \frac{M}{\mu} -1}
e^{\frac{2\,\mu^2\,R_{\rm s}^2}{\mpl^2\,\lp^2}}
\ ,
\ee
from which
\be
H_-
&\!\!\simeq\!\!&
(2\,M+\mu) \,
\frac {\Gamma\!\left(  \frac{M}{\mu}+\frac {1} {2} \right)
\left[
\Gamma\!\left(  \frac{M}{\mu}+1 \right) - 2 \, \Gamma\!\left(  \frac{M}{\mu}+\frac {3} {2} \right)
\right]} 
{\Gamma\!\left(  \frac{M}{\mu} + \frac {3} {2} \right)
\left[
\Gamma\!\left(  \frac{M}{\mu}+\frac {1} {2} \right) - 2 \, \Gamma\!\left(  \frac{M}{\mu}+ 1\right)
\right]}
\nonumber
\\
&\!\!\simeq\!\!&
2 \, \mu
\left( \frac {M} {\mu} \right)^{\frac {1} {2}}
\ .
\ee
Putting the above bounds together, we obtain
\be
M
\left[
1
+
2 \,
\left(\frac {\mu} {M}
\right)^{\frac {1} {2}}
\right]
\lesssim
\expec{\hat H}
\lesssim
M
\left(
1+
\frac{\lambda_\mu^2}
{R_{\rm s}^2}
\right)
\ ,
\label{A<H>}
\ee
which shows that $\expec{\hat H}$ cannot be smaller than the classical ADM mass and the 
(maximum) relative correction is proportional to the Compton length $\lambda_\mu=\lp\,\mpl/\mu$.
\section{Horizon wavefunction and black hole probability}
\label{A:norm}
\setcounter{equation}{0}
In Section~\ref{SS:hwf}, we continue to employ the upper bound on the spectral
decomposition to obtain the horizon wavefunction in Eq.~\eqref{psi_H} and estimate
the maximum possible correction to the gravitational radius and minimum probability
$P_{\rm BH}$.
Its normalisation is given by
\be
1
=
4\,\pi
\int\limits_{\Rh}^\infty
\left|\Psi_{\rm H}(\rh)\right|^2
\rh^2\,
\dd \rh
=
4 \, \pi \, \mathcal{N}_{\rm H}^2
\int\limits_{0}^\infty
\left(\frac{\tilde{r}_{\rm H}}{\lp}\right)^{\frac{\mpl\,\Rh}{\mu\,\lp}}
e^{-\frac {\mu\,\tilde{r}_{\rm H}\,R_{\rm s}^2} {\mpl\,\lp^3}}
\,
\left(\tilde{r}_{\rm H} + R_{\rm H} \right)^2\,\dd \tilde{r}_{\rm H}
\ ,
\ee
where $\tilde r_{\rm H}=\rh-\Rh$.
The above expression yields
\be
\mathcal{N}_{\rm H}^{-2}
&\!\!=\!\!&
\frac{8\,\pi\,\lp^9\,\mpl^3}{\mu^3\,R_{\rm s}^6}
\left(
\frac{\mpl\,\lp^2}{\mu\,R_{\rm s}^2}
\right)^{\frac{\mpl\,\Rh}{\mu\,\lp}}
\left[
\left(1+\frac{\mpl\,\Rh}{\mu\,\lp}\right)
\left(
1+\frac{\mpl\,\Rh}{2\,\mu\,\lp}
+
\frac{\mu\,\Rh\,R_{\rm s}^2}{\mpl\,\lp^3}
\right)
+
\frac{\mu^2\,\Rh^2\,R_{\rm s}^4}{2\,\lp^6\,\mpl^2}
\right]
\nonumber
\\
&&
\times\,
\Gamma\!\left(1+\frac{\mpl\,\Rh}{\mu\,\lp}\right)
\nonumber
\\
&\!\!\simeq\!\!&
\frac{4\,\pi\,\lp^3\,\mpl\,\Rh^2}{\mu\,R_{\rm s}^2}
\left(
\frac{\mpl\,\lp^2}{\mu\,R_{\rm s}^2}
\right)^{\frac{\mpl\,\Rh}{\mu\,\lp}}
\Gamma\!\left(\frac{\mpl\,\Rh}{\mu\,\lp}\right)
\ ,
\label{Nh}
\ee
where we again used $R_{\rm s}\sim\Rh\gg\lp$ in the last approximation.
\par
The expectation value of the gravitational radius is given by
\be
\bra{\Psi_{\rm H}}\hat r_{\rm H}\ket{\Psi_{\rm H}}
&\!\!=\!\!&
4\,\pi
\int\limits_0^\infty
\left|\Psi_{\rm H}(\rh)\right|^2
\rh^3\,\dd \rh
\nonumber
\\
&\!\!=\!\!&
\Rh
+
\Rh\,
\frac{\mpl^2\,\lp^2}{\mu^2\,R_{\rm s}^2}
\left(1+\frac{3\,\lp\,\mu}{\mpl\,\Rh}\right)
\nonumber
\\
&&
-
\Rh\,
\frac{
\left(1+\frac{\mpl\,\Rh}{\mu\,\lp}\right)
+\frac{\mu\,\Rh\, R_{\rm s}^2}{\lp^3\,\mpl}}
{
\left(1 + \frac{\mpl\,\Rh}{\mu\,\lp}\right)
\left(1 + \frac{\mpl\,\Rh}{2\,\lp\,\mu}\right)
+
\frac{\mu\,\Rh\,R_{\rm s}^2}{\lp^3\,\mpl}
\left(1 + \frac{\mpl\,\Rh}{\lp\,\mu}\right)
+
\frac{\mu^2\,\Rh^2\,R_{\rm s}^4}{2\,\lp^6\,\mpl^2}
}
\nonumber
\\
&\!\!\simeq\!\!&
\Rh
\left[
1
+
\frac{\lambda_\mu^2}{R_{\rm s}^2}
\left(
1
-
\frac{2\,\lp^2}{\Rh\,\lambda_\mu}
\right)
\right]
\nonumber
\\
&\!\!\simeq\!\!&
\Rh
\left(
1
+
\frac{\lambda_\mu^2}{R_{\rm s}^2}
\right)
\ ,
\label{<rh>}
\ee
with $\Rh=2 \, \gn\,M\gg\lambda_\mu$ the classical Schwarzschild radius and $\lambda_\mu\gg\lp$
the Compton length of the matter costituents.
\par
The probability density~\eqref{Phrh} for the horizon to be located on the sphere of
radius $r = \rh$ vanishes for $0\le \rh<\Rh$, else is given by
\be
\mathcal{P}_{\rm H}( r_{\rm H})
\simeq
4 \, \pi  \,
\mathcal{N}_{\rm H}^2 \,
r_{\rm H}^2 \,
\left(\frac{\rh-\Rh}{\lp}\right)^{\frac{\mpl\,\Rh}{\mu\,\lp}}
e^{-\frac {\mu\left( r_{\rm H} - R_{\rm H} \right)R_{\rm s}^2} {\mpl\,\lp^3}}
\ .
\label{Ptrh}
\ee
The probability density~\eqref{PrlessH} can be explicitly computed from the
wavefunction~\eqref{wave function in position space solution 2} with $r_i=r$
and the horizon probability density~\eqref{Ptrh},
\be
\mathcal{P}_{<}(r <\rh)
&\!\!=\!\!&
\left(4 \,\pi
\int\limits_{\Rh}^{\rh}
\left|\psi_{\rm S} (r) \right|^2 r^2 \, \dd r
\right)
\mathcal{P}_{\rm H}(\rh)
\nonumber
\\
&\!\!=\!\!&
{\rm erf} \left(  \frac {\rh} {R_{\rm s} } \right)
\mathcal{P}_{\rm H}(\rh)
\ ,
\quad
\ee
where ${\rm erf}$ denotes the error function.
The black hole probability~\eqref{Pbh} now reads
\be
P_{\rm BH}
=
\int\limits_{\Rh}^{\infty}
\mathcal{P}_{<}(r<\rh)\,
\dd  \rh
\ ,
\label{PbhC}
\ee
which however can only be computed numerically for specific values
of $\Rh$, $R_{\rm s}$ and $\mu$ (see Figs.~\ref{f:P1}-\ref{f:P3}).
\newpage
\end{document}